\documentclass[aps,twocolumn,preprintnumbers,amsmath,amssymb,superscriptaddress]{revtex4-1} 
\usepackage{amssymb}
\usepackage{amsmath}
\usepackage{xcolor}
\usepackage{hyperref}
\usepackage{braket}
\usepackage{graphicx}
\usepackage{placeins}

\begin{document}
\title{Linked Cluster Expansions via Hypergraph Decompositions}
\author{Matthias M\"uhlhauser}
\email{matthias.muehlhauser@fau.de}
\author{Kai Phillip Schmidt}
\email{kai.phillip.schmidt@fau.de}
\affiliation{Department of Physics, Friedrich-Alexander-Universit{\"a}t Erlangen-N{\"u}rnberg, Staudtstrasse 7, D-91058 Erlangen, Germany}

\begin{abstract}
We propose a hypergraph expansion which facilitates the direct treatment of quantum spin models with many-site interactions via perturbative linked cluster expansions. The main idea is to generate all relevant subclusters and sort them into equivalence classes essentially governed by hypergraph isomorphism. Concretely, a reduced K\"onig representation of the hypergraphs is used to make the equivalence relation accessible by graph isomorphism. 
During this procedure we determine the embedding factor for each equivalence class, which is used in the final resummation in order to obtain the final result.
As an instructive example we calculate the ground-state energy and a particular excitation gap of the plaquette Ising model in a transverse field on the three-dimensional cubic lattice.
\end{abstract}

\maketitle

%%%%%%%%%%%%%%%%%%%%%%%%%%%%%%%%%%%%%%%%%%%%%%%%%%%%%%%%%%%%%%%%%%%%%%%%%%%%%%%%%%%%%%%%%%%%%%%%%%%%%%%%%%%%%%
\section{Introduction}
%%%%%%%%%%%%%%%%%%%%%%%%%%%%%%%%%%%%%%%%%%%%%%%%%%%%%%%%%%%%%%%%%%%%%%%%%%%%%%%%%%%%%%%%%%%%%%%%%%%%%%%%%%%%%%
 Although graph decomposition methods have been employed already decades ago in order to obtain high-order series expansions of many-body systems \cite{Marland1981, Irving1984, He1990, Oitmaa1991, Hamer1992}, they continue to be a state of the art approach to perform calculations for quantum lattice models, for example using perturbative methods like high-temperature series expansions \cite{Hehn_2017, Singh2017}, Rayleigh-Schr\"odinger perturbation theory \cite{hehn2016series, Oitmaa2020, Oitmaa2021}, perturbative continuous unitary transformations (pCUTs) \cite{Knetter_2000,  Knetter2003, Hong2010, Coester2016, adelhardt2020, Schmoll2020} or non-perturbative numerical tools like exact diagonalization \cite{Ixert2016, Rigol_2006, Schaefer_2020}, density matrix renormalization group \cite{Stoudenmire2014}, and graph-based continuous unitary transformations \cite{Yang2011,Ixert2014, Coester2015} in the context of numerical linked cluster expansions.
 Historically the graph decomposition methods for high-order series expansions were exclusively applicable to extensive quantities like ground-state energies \cite{Marland1981,  Irving1984} or had to take into account also disconnected graphs \cite{He1990, Gelfand1996}. It took some time until it was clear how to set up linked cluster expansions, where only connected graphs have to be considered, for non-extensive quantities like the excitation gap \cite{Gelfand1996}, two-particle properties \cite{Trebst2000, Zheng2001, Knetter2000Bound} or multi-particle spectral densities \cite{Schmidt2001, Knetter2001Fractional, Hoermann2018}. Detailed reviews on high-order series expansions using graph decomposition techniques can be found in \cite{Gelfand2000, oitmaa_hamer_zheng_2006}.
 While for pairwise interactions it is straightforward to represent the involved clusters in terms of graphs, where the vertices correspond to sites and the edges represent interactions between the sites \cite{Joshi2015, Hehn_2017}, this is usually not the case when interactions join multiple sites.
 However, some problem specific approaches to treat such many-site interactions in topologically ordered phases can be found in the literature \cite{Schulz2013, Schulz2014}.
 \par 
 Many-site interactions play an important role for several aspects in quantum many-body systems. In statistical physics, interesting critical properties arise due to three-body interactions like in the Baxter-Wu model \cite{Baxter-Wu1973, Capponi2014} or quantum phase transitions with fractal properties \cite{zhou2021}. In topological systems, there exist a collection of exactly solvable models with exotic physical properties like topological or fracton order which consist of commuting many-site operators - so called stabilizer codes \cite{Gottesman1997}. Famous examples are Kitaev's toric code \cite{Kitaev2003} and its three-dimensional generalization \cite{Hamma2005, Nussinov2008, Reiss2019}, color codes \cite{Kargarian2009, Katzgraber2009}, and string-net models \cite{Levin2005}
  displaying long-range entangled ground states with topological order or three-dimensional fracton systems like the X-Cube model, the checkerboard model or Haah's cubic code \cite{Haah_2011, Vijay2016}. Another class of stabilizer codes with multi-site interactions are cluster-state Hamiltonians in the context of measurement-based quantum computation \cite{Raussendorf2001, Briegel2001, Orus2013}. The exact solvability of these stabilizer codes has been the starting point for many fundamental investigations of the physical properties of such systems, e.g.~their quantum robustness and associated quantum critical properties. High-order series expansions represent one important tool to study these questions \cite{Dusuel2011, Kalis2012, Jahromi2013, Schulz2013, Schulz2014, Muehlhauser2020}. In condensed matter physics, multi-site interactions are known to arise in strongly correlated Mott insulators like Hubbard systems in the limit of strong interactions \cite{MacDonald1988, Reischl2004, Yang2010, Yang2012}. In unfrustrated materials like the undoped high-$T_{\rm c}$ superconductors or corresponding quasi one-dimensional cuprate ladders, four-spin interactions are needed for a quantitative description of the magnetic properties \cite{Windt2001, Katanin2003, Schmidt2005, Schmidt2005Spectral, Notbohm2007}. In frustrated Mott insulators like the Hubbard model on the triangular lattice, four-spin interactions are expected to trigger exotic quantum spin liquid phases in the Mott insulating regime \cite{Motrunich2005, Sheng2009, Yang2010, Block2011, Boos2020, Szasz2020}.

 %Such many-site interactions arise in a lot of models which are of theoretical interest \cite{Baxter-Wu1973, Xu2004, Orus2013, Jahromi2013, Schulz2013, Capponi2014, zhou2021}. 
%One of the most famous models which hosts many-site interactions is Kitaev's toric code \cite{Kitaev2003}. But
%there are many other models which feature many-site interactions e.g. the three-dimensional generalization of the toric code \cite{Hamma2005, Nussinov2008} or fracton models like the X-Cube model, the Checkerboard model or Haah's cubic code \cite{Haah_2011, Vijay2016}. 
 
As a consequence, a general framework for performing linked cluster expansions in quantum many-body systems with multi-site interactions is highly desirable. From a conceptional point of view hypergraphs are a natural generalization of graphs to describe more complex relationships, where edges can connect more than two vertices \cite{Berge1973, Zykov1974, Berge1989, Bretto2013, ouvrard2020hypergraphs}. Indeed, they have also been employed to represent molecular structures \cite{Konstantinova1995, Konstantinova2001}. Additionally, the isomorphism problem for hypergraphs can be solved using conventional graph isomorphism as hypergraphs are uniquely defined by their K\"onig representation \cite{Zykov1974, mckay1981practical}. 

In this work we describe how these concepts can be used to perform a graph decomposition for many-site interactions.  While in \cite{Muehlhauser2020} we applied a similar technique to study the X-Cube model in a magnetic field, we used the same approach as presented in this work to investigate the competition between the X-Cube and the three dimensional toric code in \cite{Muehlhauser2021} without, however, describing the technical aspects. 
%We emphasize, that while we did employ a slightly different technique in \cite{Muehlhauser2020} we already used an earlier version of the presented method to investigate the competition between the X-Cube and the 3D-Toric code \cite{Muehlhauser2021}. 
As a representative example, we apply our scheme to the plaquette Ising model \cite{Savvidy1995, Johnston2017}, in a transverse magnetic field on the three-dimensional cubic lattice \cite{Nandkishore2019}.

%Prior to this we briefly review the idea of perturbative linked cluster expansions in the context of pCUTs which has already been described in \cite{coester2015optimizing}.
The paper is organized as follows. In Sec.~\ref{sect::pCUT} we describe perturbative linked cluster expansions in the context of pCUTs \cite{Knetter_2000, Knetter2003}, as introduced in \cite{coester2015optimizing}, with a focus on multi-site interactions. The general aspects of our scheme are contained in Sec.~\ref{sect::scheme} and it is applied to the plaquette Ising model in a transverse field on the cubic lattice in Sec.~\ref{sect::application}. Final conclusions are drawn in Sec.~\ref{sect::conclusion}.

%%%%%%%%%%%%%%%%%%%%%%%%%%%%%%%%%%%%%%%%%%%%%%%%%%%%%%%%%%%%%%%%%%%%%%%%%%%%%%%%%%%%%%%%%%%%%%%%%%%%%%%%%%%%%%
\section{Linked cluster expansions using pCUTs}
\label{sect::pCUT}
%%%%%%%%%%%%%%%%%%%%%%%%%%%%%%%%%%%%%%%%%%%%%%%%%%%%%%%%%%%%%%%%%%%%%%%%%%%%%%%%%%%%%%%%%%%%%%%%%%%%%%%%%%%%%%
In this section we review pCUTs in the context of linked cluster expansions based on \cite{coester2015optimizing} focusing on multi-site interactions. For a more detailed introduction to pCUTs we refer to \cite{Knetter_2000, Knetter2003}. We consider a lattice Hamiltonian of the form
\begin{equation} H = H_0 + \lambda V \, .\end{equation} 
Note that one can easily introduce several perturbation parameters $\lambda_1 \ldots \lambda_n $, but we will restrict ourselves to a single perturbation parameter $\lambda$ for the sake of simplicity.
For the unperturbed part of the Hamiltonian we make the same assumptions as \cite{coester2015optimizing}, i.e. that it can be written as
\begin{equation} H_0 = E_0 + Q = E_0 + \sum_i n_i \, , \end{equation}
where $Q$ is the global particle number-operator, $n_i$ are local particle-number operators at the site $i$, and $E_0$ is a constant. In the example presented in Sec.~\ref{sect::application} the local degrees of freedom on sites are one type of hardcore boson. However, a site can in principle host systems which feature several degrees of freedom, which are then commonly called supersites.
The interaction $V$ contains interactions between these sites.
In contrast to the discussion in \cite{coester2015optimizing} we do not restrict ourselves to two-site interactions. 
 \par
The effective Hamiltonian resulting from pCUTs is given as \cite{Knetter_2000, Knetter2003}
\begin{equation} 
 H_{\text{eff}} = H_0 + \sum_{k=1}^\infty \lambda^k  \sum_{ \substack{\vec{m}, \vert \vec{m} \vert = k  \\ \sum_i m_i = 0}}   \mathcal{C}(\vec{m})  T(\vec{m}) \,   
\end{equation}
with
\begin{align}
 \vec{m} &= (m_1, \ldots, m_k) \, ,\\
 \vert \vec{m} \vert &= \text{dim}(\vec{m}) \, ,\\
 T(\vec{m}) &= T_{m_1} \ldots T_{m_k} 
\end{align}
and the $T_n$-operators, which create $n$ quasiparticles
\begin{equation}
	[Q, T_n] = n T_n \, .
\end{equation}
The most important property of $H_{\text{eff}}$ is $[H_{\text{eff}},Q]=0$ reflecting the quasi-particle conservation. The effective Hamiltonian $H_{\text{eff}}$ is therefore block-diagonal in the number of quasi-particles so that the complicated quantum many-body problem is mapped to an effective few-body problem by pCUTs. This step is model-independent. The model-dependent step of pCUTs is to normal order $H_{\text{eff}}$ in order to extract physical properties like ground-state energies, one-quasi-particle hopping amplitudes or two-quasi-particle interactions. This step is most efficiently done by calculating matrix elements of $H_{\text{eff}}$ on finite clusters exploiting the linked-cluster property of pCUTs.

Obviously, every matrix element of $H_{\text{eff}}$ can be straightforwardly evaluated on a single appropriately chosen cluster, but especially for high-order calculations these clusters tend to become very large making the evaluation of the effective Hamiltonian inefficient \cite{Dusuel2010}. 
At this stage a graph decomposition can be employed to reduce the computational demands. 
A first step is to realize that the $T_n$-operators can be decomposed into local operators \cite{coester2015optimizing}
\[ T_n = \sum_{b} \tau_{n,b}\, . \] 
These local operators usually act only on a small number of sites which are joined by a bond $b$, where they 
create $n$ quasiparticles. Recall that in order to model many-particle interactions we allow bonds to connect multiple sites.
Using this notion, the effective Hamiltonian is rewritten as
\begin{equation} H_{\text{eff}} = H_0 + \sum_{k=1}^\infty \lambda^k  \sum_{ \substack{\vec{m}, \vert \vec{m} \vert = k \\ \sum_i m_i = 0} } \sum_{\vec{b}, \vert \vec{b} \vert = k }  \mathcal{C}(\vec{m}) \tau(\vec{m}, \vec{b}) \,  \end{equation}
with
\begin{align}
\vec{b} &= (b_1, \ldots b_k) \\
\vert \vec{b} \vert &= \text{dim}(\vec{b}) \\
\tau(\vec{m}, \vec{b}) &= \tau_{m_1,b_1} \ldots \tau_{m_k,b_k} \, .
\end{align}

Rearranging the sums we can extract the following expression for any bond sequence $\vec{b}$
\begin{equation} V_{\vec{b}} =    \lambda^{\vert \vec{b} \vert }  \sum_{ \substack{\vec{m}, \vert \vec{m} \vert = \vert \vec{b} \vert \\ \sum_i m_i = 0 }} \mathcal{C}(\vec{m})  \tau(\vec{m}, \vec{b}) \, , \end{equation}
and the effective Hamiltonian becomes
\begin{equation}
H_{\text{eff}}= H_0 +  \sum_{\vec{b}} V_{\vec{b}}  \, . \label{eq::H_divided} \end{equation}
Due to the linked cluster property of pCUT the contributions on disconnected 
subclusters $\mathcal{S}$ vanish \cite{kamfor2013robustness, coester2015optimizing}.
This linked cluster property can be directly understood from the commutator structure of the effective Hamiltonian \cite{Dusuel2010}. 
A bond sequence $\vec{b}$ can be naturally associated to the subcluster of the full
system, which exactly contains the bonds which are included in $\vec{b}$.
Accordingly, the sum over bond sequences becomes a sum over connected subclusters $\mathcal{S}$

\begin{align}
H_{\text{eff}} &= H_0 +  \sum_{\mathcal{S}} \sum_{\vec{b} \mapsto \mathcal{S}} V_{\vec{b}} \\
&= H_0 +  \sum_{\mathcal{S}} V_{\mathcal{S}}\, ,
\end{align}
where the subscript $ \vec{b} \mapsto \mathcal{S}$ means that the sum runs over all bond sequences $\vec{b}$ associated with the subcluster $\mathcal{S}$. Note that typically many bond sequences are associated to the same subcluster $\mathcal{S}$, since individual bonds can be touched multiple times and in different orders.
This implies that evaluating $V_{\mathcal{S}}$ only processes which involve all bonds in $\mathcal{S}$
have to be taken into account.
\par 
A crucial next step is to realize, that we do not have to evaluate $V_{\mathcal{S}}$ on every subcluster, but only once for every equivalence class of subclusters.
The actual equivalence relation depends on the matrix element $\bra{\Phi} H_{\text{eff}} \ket{\Psi}$ for which the effective Hamiltonian is evaluated. 
In case $\ket{\Phi}, \ket{\Psi}$ are the bare vacuum state ($n_i= 0 \; \forall i$) it is sufficient to identify isomorphic clusters. This is not the case for excited states, as the positions of excitations in the states $\ket{\Phi}, \ket{\Psi}$ matter.
Generically, a matrix element of the effective Hamiltonian can be written as
 \begin{equation}
 \begin{aligned}
 \bra{\Phi} H_{\text{eff}} \ket{\Psi} &= \bra{\Phi} H_0 \ket{\Psi} +  \sum_{\mathcal{E}} N_{\mathcal{E}}   \bra{\Phi} V_{\mathcal{S} \in \mathcal{E}} \ket{\Psi}   \\
 & =  \bra{\Phi} H_0 \ket{\Psi} + \sum_\mathcal{E} N_{\mathcal{E}} V_{\mathcal{E}} \, ,
 \end{aligned}
 \end{equation}
 where $\mathcal{S} \in \mathcal{E}$ labels a representative subcluster which is contained in the equivalence class $\mathcal{E}$, and $N_{\mathcal{E}}$ is the number of subclusters contained in $\mathcal{E}$.
 Obviously, the matrix element $V_{\mathcal{E}}$ can be decomposed, such that 
 one factor is evaluated only on the subcluster, and the other is just the scalar product of the states on the remaining Hilbert space. 
Accordingly, the effective Hamiltonian is evaluated only on one subcluster per equivalence class, effectively exploiting symmetries of the model and the lattice.
 The price to pay is, that we have to find all equivalence classes and determine the embedding factors
$N_\mathcal{E}$ \cite{Dusuel2010}, which in the presented approach are essentially obtained by counting the relevant subclusters, i.e.~typically one does not have to count all subclusters but one can use process-specific properties like translational invariance for the ground-state energy or the finite extension of linked quantum fluctuations in case of hopping amplitudes.
 \par
 For the calculation of ground-state expectation values, this graph decomposition approach also works with other methods to evaluate the matrix elements $V_{\mathcal{E}}$ on graphs like Takahashi's perturbation theory \cite{Takahashi1977, Klagges2012}, L\"owdin's partitioning technique \cite{Loewdin1962, Kalis2012}, or matrix perturbation theory \cite{oitmaa_hamer_zheng_2006}, because the series are unique on any graph and therefore do not depend on the employed method.
 
%%%%%%%%%%%%%%%%%%%%%%%%%%%%%%%%%%%%%%%%%%%%%%%%%%%%%%%%%%%%%%%%%%%%%%%%%%%%%%%%%%%%%%%%%%%%%%%%%%%%%%%%%%%%%%
\section{Scheme}
\label{sect::scheme}
%%%%%%%%%%%%%%%%%%%%%%%%%%%%%%%%%%%%%%%%%%%%%%%%%%%%%%%%%%%%%%%%%%%%%%%%%%%%%%%%%%%%%%%%%%%%%%%%%%%%%%%%%%%%%%
The presented scheme consists of two main parts: generating all relevant subclusters and sorting them into equivalence classes based on hypergraph isomorphism. Actually we are doing this on the fly, i.e., whenever we discover a relevant subcluster we check whether we already found an equivalent cluster before. In this way we accumulate a set of equivalence classes while we can also keep track of the corresponding embedding factors. For the sake of simplicity we assume that all bonds in the lattice are symmetry equivalent, and undirected, i.e., the local perturbation
 acts on all involved sites in the same way. We address in App.~\ref{appendix::complicatedmodels} how this scheme can be applied without these assumptions. We further note that a bond does never contain the same site more than once.\par
It is known that for conventional graphs first generating all relevant isomorphism classes and then calculating the respective embedding factors is more efficient \cite{Gelfand2000}. The presented scheme is no exception. However, our choice is justified taking into account that the generation of hypergraph isomorphism classes appears more challenging compared to the generation of graph isomorphism classes, although it is actually covered in the literature \cite{mckay1998isomorph}. Even if the problem of finding all relevant hypergraphs was solved, we would still expect the calculation of the embedding factors to be quite time-consuming.

%%%%%%%%%%%%%%%%%%%%%%%%%%%%%%%%%%%%%%%%%%%%%%%%%%%%%%%%%%%%%%%%%%
\subsection{Generation of relevant subclusters}  
\label{sect::generation}
%%%%%%%%%%%%%%%%%%%%%%%%%%%%%%%%%%%%%%%%%%%%%%%%%%%%%%%%%%%%%%%%%%
The generation of subclusters is based on an algorithm presented in \cite{Ruecker_2000}, which enumerates all connected subgraphs of a given (conventional) graph. 
Although this algorithm is described in the literature for conventional graphs the adaption for multi-site interactions described below is straightforward, as the concept to identify a connected subcluster by the set of its bonds is independent of the bond cardinality.\par
The algorithm starts from a single bond $b_1$ and builds connected subclusters by adding further bonds in a depth-first manner \cite{Ruecker_2001}.
The candidates to be appended are simply the bonds adjacent to the actual subcluster.
If a feasible candidate is found, this bond is appended to the subcluster and the algorithm continues with the extended subcluster.
The algorithm backtracks if no feasible candidate is found, i.e., it removes the last appended bond and marks it as forbidden. The bond remains forbidden until the algorithm backtracks one step further ensuring that every connected subcluster containing $b_1$ is found once and only once \cite{Ruecker_2000}. 
Marking $b_1$ as forbidden and restarting the search from another bond $b_2$ will yield all connected subclusters which contain $b_2$ but do not contain $b_1$. 
So in order to find all connected subclusters the search has to be repeated from all possible starting bonds ignoring all the previous starting bonds during the search \cite{Ruecker_2000}.
\par
Obviously, this algorithm can be easily modified to produce all connected subclusters up to a given number of bonds by simply forcing it to backtrack when the maximum size is reached \cite{Ruecker_2000}. 
Additionally, the structure of the search tree makes it easy to incorporate heuristics \cite{oitmaa_hamer_zheng_2006} to discard unfruitful branches in advance. \par
For our purpose it is very useful that the algorithm actually allows to generate only the clusters, which intersect with a given set of bonds.
For the calculation of hopping amplitudes this helps to only consider clusters which host fluctuations different from vacuum fluctuations, whereas for the calculation of the ground-state energy per bond, this is useful, as we can restrict the set of generated clusters to the clusters which actually contain this bond. \par
Interestingly, this algorithm has been combined with highly discriminating graph invariants to find all connected non-isomorphic subgraphs of a given graph \cite{Ruecker_2001}. However, while this approach yields exact results in some domains, it is only an approximate solution, as the proposed graph invariants were not complete, i.e., there exist non-isomorphic graphs, which have the same invariants \cite{Ruecker_2001, Ruecker_2002}. \par

%%%%%%%%%%%%%%%%%%%%%%%%%%%%%%%%%%%%%%%%%%%%%%%%%%%%%%%%%%%%%%%%%%
\subsection{Sorting subclusters into equivalence classes}
\begin{figure}
\includegraphics[width = 0.4 \textwidth]{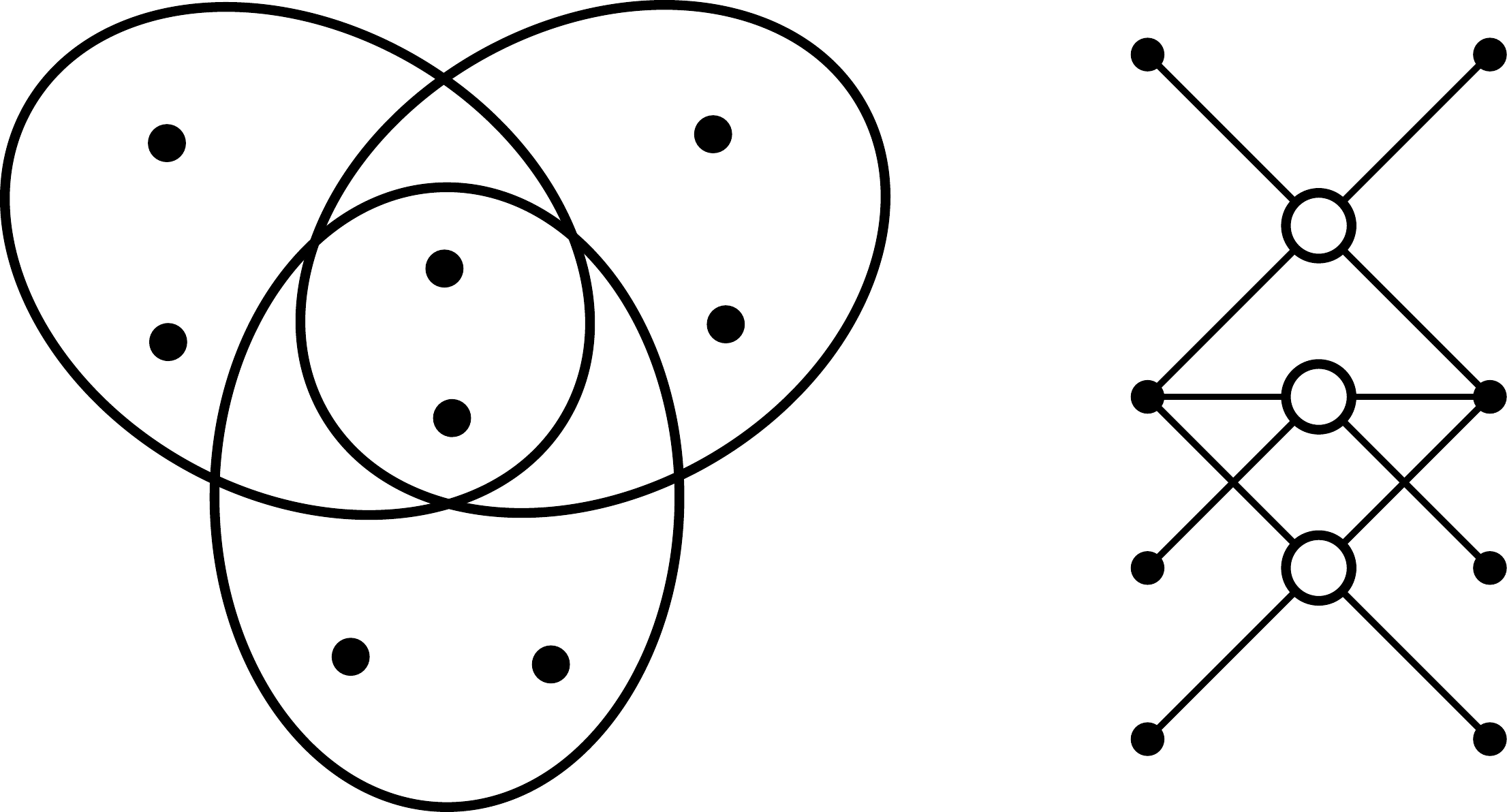}
\caption{An exemplary hypergraph (left) and its K{\"o}nig representation (right). On the left side the black dots correspond to the vertices and the ellipses correspond to the hyperedges linking the included vertices. On the right side the filled vertices correspond to the vertex partition, while the unfilled vertices correspond to the edge-partition i.e., they represent the hyperedges.
	Vertices from the two partitions are connected if and only if, the respective vertex is contained in the corresponding hyperedge \cite{Zykov1974}. 
	\label{fig::koeniggraph}
 }
\end{figure}

%%%%%%%%%%%%%%%%%%%%%%%%%%%%%%%%%%%%%%%%%%%%%%%%%%%%%%%%%%%%%%%%%%
\label{subsect::sorting}
In the presented approach subclusters are sorted into equivalence classes based on hypergraph isomorphism.
To this end subclusters are associated with the K{\"o}nig representation of the corresponding hypergraph. 
The K{\"o}nig representation of a hypergraph is a bipartite graph, where one partition represents the vertices of the hypergraph while the other partition represents the hyperedges \cite{Zykov1974}. Whenever a vertex is contained in a hyperedge, an edge connects the corresponding vertices in the bipartite graph. Taking into account the bipartition of the vertex set, this graph represents the hypergraph unambiguously \cite{Zykov1974, Konstantinova1995}.
Accordingly, this representation enables us to use conventional graph isomorphism to discriminate isomorphism classes of subclusters. An example for the K{\"o}nig representation of a hypergraph is shown in Fig. \ref{fig::koeniggraph} \footnote{Remarkably, an illustration in Ref.~\cite{Schulz2013phd} is actually showing a K{\"o}nig graph, but then a different slightly more problem specific approach is used to distinguish equivalence classes of clusters.}. 
Note that one can easily distinguish individual sites by coloring the corresponding vertices in the König representation. With this adaption it is also possible to include information about the states involved in a matrix element and so also in this case one can distinguish equivalence classes by graph isomorphism.
\par
In practice, we do not use the full K{\"o}nig representation, but a reduced representation, which exploits
that the number of sites in all bonds is equal, since we have assumed all bonds to be symmetry equivalent. So vertices of degree one in the vertex partition and adjacent edges can be ignored, if they do not carry any additional information. As a consequence, for calculations of ground-state energies all such vertices can typically be ignored, while for the calculation of hopping amplitudes the distinguished vertices, necessary to encode the initial and the final state of the actual matrix element, have to be kept. \par
  We further perform isomorphism checks with the RI algorithm \cite{Bonnici2013subgraph, Bonnici2016variable} among the resulting graphs. Since such checks are typically rather expensive, we use a combination of graph invariants in order to apply these checks only to graphs which have the same invariants. \par
  While just sorting the relevant subclusters into isomorphism classes is sufficient for the calculation of hopping amplitudes, one still has to apply some corrections for the calculation of the ground-state energy when exploiting symmetries like translations. In a conventional graph decomposition an edge of the graph is typically fixed to a certain bond in the lattice in order to remove redundancies due to existing symmetries like translations. In our approach, this corresponds to the requirement that this bond in the lattice has always a canonical location in the subclusters. If this requirement is not fulfilled on a subcluster, it is not counted as a valid embedding. We will not further elaborate on the details here, but refer to the example
  presented in Sec.~\ref{sect::application}, where we also comment on how to keep track of the embedding factors.  
 
 \FloatBarrier

%%%%%%%%%%%%%%%%%%%%%%%%%%%%%%%%%%%%%%%%%%%%%%%%%%%%%%%%%%%%%%%%%%%%%%%%%%%%%%%%%%%%%%%%%%%%%%%%%%%%%%%%%%%%%%
\section{Application}
\label{sect::application}
%%%%%%%%%%%%%%%%%%%%%%%%%%%%%%%%%%%%%%%%%%%%%%%%%%%%%%%%%%%%%%%%%%%%%%%%%%%%%%%%%%%%%%%%%%%%%%%%%%%%%%%%%%%%%%
In this section we treat the plaquette Ising model \cite{Savvidy1994, Johnston2017} in a transverse magnetic field on the cubic lattice \cite{Nandkishore2019} as a representative example. The Hamiltonian is given as 
\begin{equation}  H = - \sum_i \sigma^z_i - \lambda \sum_{\square}  \sigma_{i_1}^x \sigma_{i_2}^x \sigma_{i_3}^x \sigma_{i_4}^x \, , \label{eq::application}\end{equation} 
where $\square$ indicates that the sum runs over all plaquettes in the lattice, and the indices $i_1, \ldots i_4$ refer to the four sites at the corners of the respective plaquette.
Interestingly, this model is dual to the relevant low-energy sector of the fractonic X-Cube model in a $z$-field \cite{Vijay2016, Nandkishore2019}. For this case we already obtained series expansions for the ground-state energy and excitations gaps of one- and two-particle excitations \cite{Muehlhauser2020}. For the ground-state energy we also compared to quantum Monte Carlo simulations received from Trithep Devakul \cite{Devakul2018} which are also displayed in \cite{Muehlhauser2020}. In the latter work we also made the first steps using graph decomposition techniques for fracton models. In the following we aim at treating this model about the high-field limit $\lambda \ll 1 $ using high-order linked-cluster expansions. We will start by discussing how to calculate the ground-state energy using the presented approach in large detail. Afterwards we will briefly comment on how this scheme can be adapated to calculate hopping elements of quasiparticle excitations.
%%%%%%%%%%%%%%%%%%%%%%%%%%%%%%%%%%%%%%%%%%%%%%%%%%%%%%%%%%%%%%%%%%
\subsection{Ground-state energy}
%%%%%%%%%%%%%%%%%%%%%%%%%%%%%%%%%%%%%%%%%%%%%%%%%%%%%%%%%%%%%%%%%%

The ground state of the unperturbed system $\lambda=0$ is given by the polarized state
\begin{equation} \ket{\mathrm{GS}} = \ket{\uparrow \ldots \uparrow} \, .\end{equation} 
The action of each four-spin interaction in the perturbation flips the four $\sigma^z$-eigenvalues at the four sites of a plaquette.
\par

\subsubsection{Mapping clusters to graphs}
\begin{figure}[h]
	\centering
	\includegraphics[width=0.2 \textwidth]{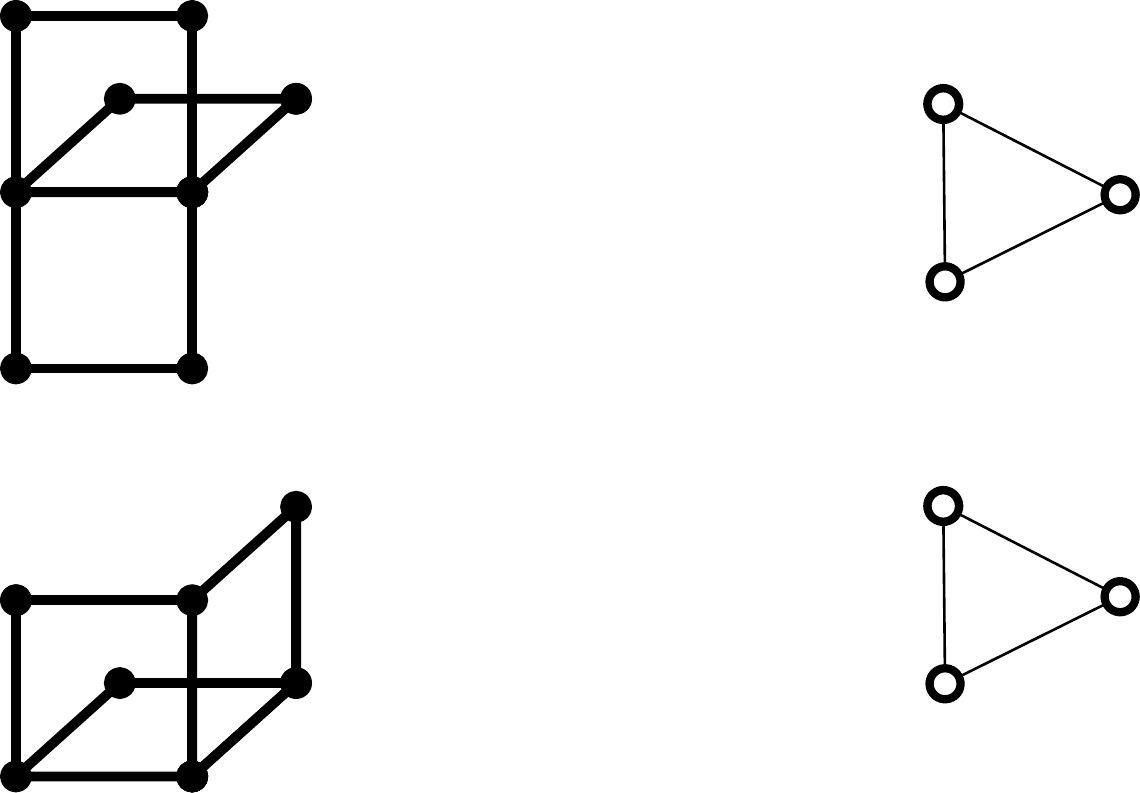}
	\caption{
		On the left side two subclusters are given, while on the right side their representation using the 
		naive approach presented in the text is illustrated. Empty circles represent plaquettes and they are connected because they have two common sites (another edge color could be used if they would share one site only).
		In this naive approach the above sublusters would be errorneously classified as equivalent.}
	\label{fig:naiveapproach}
\end{figure}

\begin{figure}[h]
	\includegraphics[width = 0.3 \textwidth]{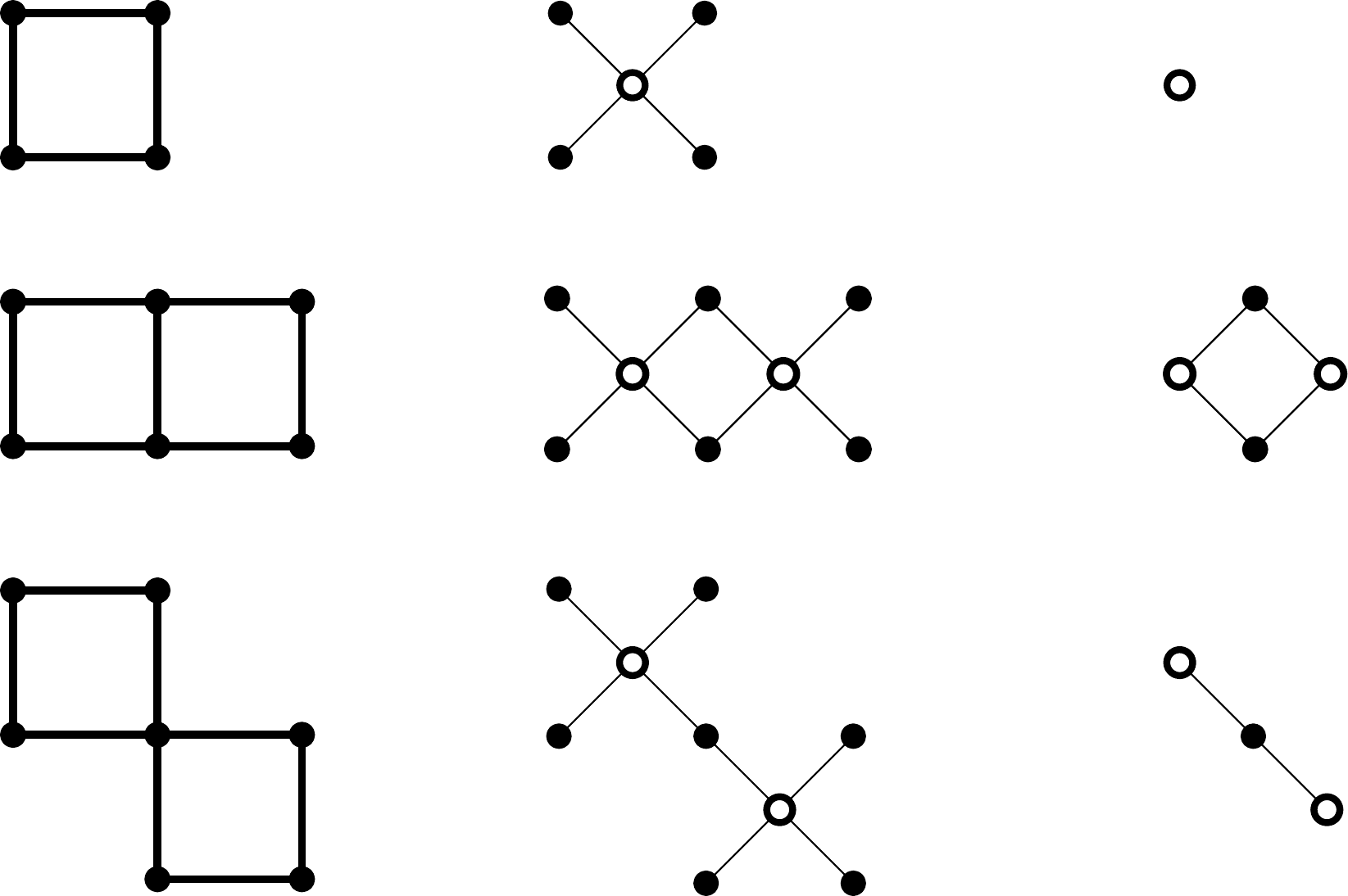}	
	\caption{Different isomorphism classes of subclusters up to two plaquettes. In the left column  subclusters are given as found in the actual lattice i.e.~every plaquette corresponds to a bond, which links the four included sites. The central column is the K{\"o}nig representation of the corresponding hypergraph. The right column gives a reduced K{\"o}nig representation, which we can use here, because we know that all bonds have cardinality four, and that we do not have to distinguish the sites, because the ground state is simply a polarized state.}
	\label{fig:graphs_upto_2}
\end{figure}

\begin{figure}[htbp]
	\includegraphics[width = 0.3 \textwidth]{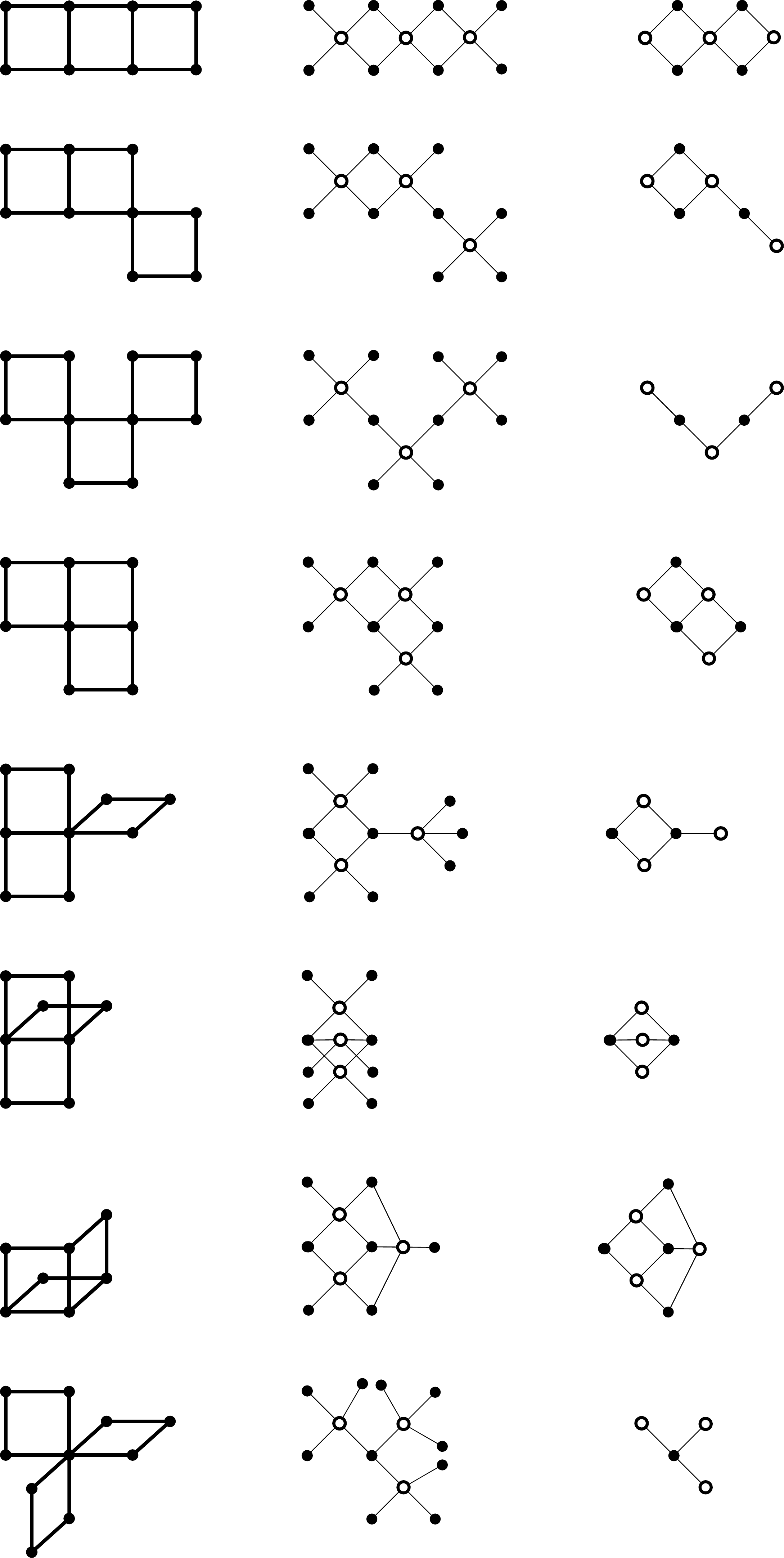}	
	\caption{Different isomorphism classes of subclusters with three plaquettes. The columns are explained as in Fig.~\ref{fig:graphs_upto_2}. 
	In the left column subclusters are given as found in the actual lattice i.e.~every plaquette corresponds to a bond, which links the four included sites. The central column is the K{\"o}nig representation of the corresponding hypergraph. The right column gives a reduced K{\"o}nig representation, which we can use here, because we know that all bonds have cardinality four, and that we do not have to distinguish the sites, because the ground state is simply a polarized state.
	In this order the involved graphs become more complicated and already at this stage a naive approach starts to fail, as shown in Fig.~\ref{fig:naiveapproach}.}
	\label{fig:graphs_3}
	
\end{figure}

First, we observe that bonds always connect four sites. Hence, the common interpretation of bonds as edges of a graph is hardly possible. At this point it is tentative to search for straightforward mappings. For the present problem a naive approach would consist in representing each plaquette by a vertex, and connecting pairs of these vertices by colored edges where the edge color encodes the amount of sites shared by the respective plaquettes.
This naive mapping to conventional graphs already fails for clusters involving three plaquettes (see Fig. \ref{fig:naiveapproach} for a counterexample). 
%\kainew{Interestingly, in two dimensions, this mapping works for the same Hamiltonian on the square lattice.} 
Obviously, if such a simple mapping is sufficiently discriminating for the problem at hand, it will very likely outperform the presented method. However, it is not always easy to find such a mapping and there can be some subtle pitfalls.  
Instead, we interpret the plaquettes as edges of a hypergraph, and use the K{\"o}nig representation of this hypergraph to represent the isomorphism class of the subcluster. 
The relevant isomorphism classes up to two plaquettes are given in Fig.~\ref{fig:graphs_upto_2}, while the classes for three plaquettes are illustrated in Fig.~\ref{fig:graphs_3}, showing that the presented approach correctly discriminates the two subclusters in Fig.~\ref{fig:naiveapproach}. 
Clearly, using the full K{\"o}nig graph will be very inefficient, especially in cases where bonds connect many sites. However, we can easily reduce the representation, by simply omitting vertices from the vertex partition which have degree one and the adjacent edges.
In Figs.~\ref{fig:graphs_upto_2} and \ref{fig:graphs_3} it can be seen in the right column how this reduced representation simplifies the involved graphs.\par

\subsubsection{Identifying non-contributing clusters}
\label{subsubsect::non-contributing}
Another important consideration is to identify and discard subclusters which cannot contribute as early as possible.
Heuristics to discard non-contributing clusters can be found in \cite{He1990, oitmaa_hamer_zheng_2006, coester2011series, Joshi2015}. 
As appropriate variants of these heuristics are also very useful for the problem at hand, we try to motivate two of them in the context of \eqref{eq::application}.
First, we distinguish between transitive and intransitive heuristics. 
While an intransitive heuristic only states that the cluster in question cannot contribute, a transitive heuristic also concerns clusters which are obtained by expanding this cluster during the search process up to a given order.
The authors of \cite{oitmaa_hamer_zheng_2006} describe a transitive heuristic which filters out graphs based 
on the degrees of the involved vertices.
The rationale behind this technique is the following: As the action of the perturbation at a bond flips all spin-eigenvalues affected by that bond, the perturbation is required to act an even amount of times on all sites in order to reobtain the vacuum state. 
Recalling that the perturbation has to act at least once on every bond in the subcluster, we see that for any site corresponding to a vertex of odd degree (odd-degree site), some bond adjacent to the respective site has to be touched multiple times by the perturbation. Of course, there can be multiple odd-degree sites in a single bond, but in the end this number is limited by the maximal bond cardinality in the system, which is four for the example at hand. So we find the condition
\[ N_{\text{odd}} \leq 4  \cdot (O_\text{pert}  -  N_{\text{bonds}}) \, , \]
where $N_{\text{odd}}$ is the number of odd-degree vertices in the vertex partition of the full K{\"o}nig graph, $N_{\text{bonds}}$ is the number of bonds (i.e.~the number of vertices in the edge-partition) and $O_\text{pert} $ is the desired perturbation order. If this condition is false, the investigated subcluster and all subclusters which are found by expanding it, can be discarded. 
This condition is very useful, especially if it is checked before the actual graph representation is calculated. Furthermore, it is also suited as a preliminary check for more involved techniques due to its simplicity. \par
While searching and classifying subclusters, we do have the information about the embedding of the subcluster into the whole system. Hence, we can determine, whether $(O_\text{pert}  -  N_{\text{bonds}})$ of all bonds in the system, can cover all odd-degree sites in the subcluster.
For any pair of sites in the system, we store from the beginning of the calculation, whether they do share a common bond or not. We now try to find a lower bound on the number of bonds needed to cover all odd-degree sites. Accordingly, we search the minimum partition, i.e.~the one with the smallest number of parts, of the odd-degree sites of the investigated subcluster such that sites which do not share a bond are in distinct parts. It is well known that such problems can be solved using graph coloring \cite{Lewis2016}. To this end we construct an auxiliary graph where we represent any odd-degree site of the investigated subcluster by a vertex, and add an edge whenever two odd-degree sites cannot be contained in the same bond.
If the constructed graph is not $r$-colorable, with  $r = (O_\text{pert}  -  N_{\text{bonds}})$, the investigated subcluster and all subclusters which are found by expanding it, can be discarded. Note that also this condition can be checked before the actual graph representation of the subcluster is calculated. An example and further details of this technique are given in appendix \ref{appendix::graphcoloring} \par
Furthermore, we also used an intransitive heuristic, which is also motivated by the fact that every site has to be touched an even number of times in order to recast the vacuum. 
The technique consists in finding the minimum number of bonds, which have to be duplicated in order to eliminate all odd-degree sites in a subcluster which represents an isomorphism class \cite{He1990}. 
If the number of bonds of the appropriately extended cluster exceeds the maximum perturbation order, the entire isomorphism class is not relevant for the calculation.
\par
We conclude with the general remark, that while in principle a small set of routines can be employed to deal with a variety of models, the heuristics are still strongly problem dependent and there are also many models where it is not possible to reduce the size of the search tree or the number of clusters which have to be considered during the final evaluation with such routines. \par

	\begin{figure}[htbp]
	\label{fig::orbit}
	\begin{center}
	\includegraphics[width = 0.48 \textwidth]{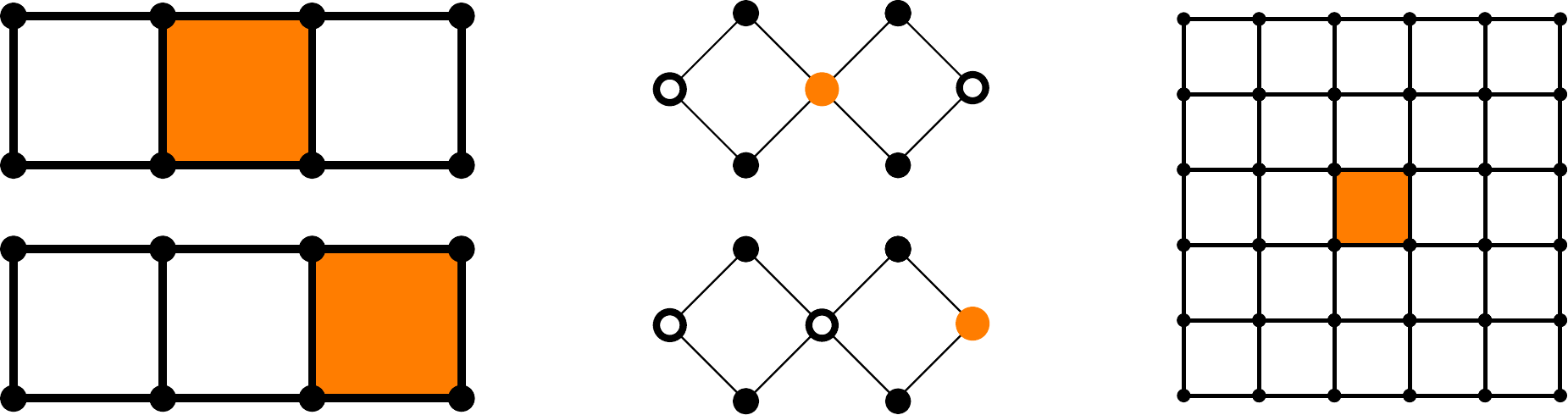}
	\end{center}
	\caption{In the left column we show two clusters which would be obtained by starting the search process on the lattice in the right column from the orange bond (plaquette). 
	In the central column the reduced K\"onig representations are shown. The orange vertices indicate the position of the first vertex for the two isomorphic graphs. When enforcing the canonical location of the first vertex, only one of the subclusters has to be considered. If one chooses the first subcluster, then the location of the first vertex in the graph is unique. In contrast, for the lower subcluster, there are two valid locations for the first vertex due to the reflection symmetry. This ambiguity has to be corrected by dividing the number of embeddings by 2. Our approach can therefore be seen as a bottom-up version of the embedding process described in \cite{oitmaa_hamer_zheng_2006}, where an edge in the graph is fixed to a bond in the lattice during the embedding process.
	}
	
\end{figure}
	
\subsubsection{Putting the pieces together}

The vacuum energy is an extensive quantity so we aim to calculate the vacuum energy per plaquette.
As a starting point we recall from Sec.~\ref{sect::generation} that we could find all connected subclusters up to a given number of bonds which include a given starting bond. For further explanations the vertex representing the starting bond will be referred to as the first vertex. In order to find the number of subclusters per plaquette, the first vertex must always have a canonical location in the subclusters. If this requirement is not fulfilled on a subcluster, it is not counted as a valid embedding. Enforcing the canonical location of the first vertex corresponds to fixing an hyperedge of the respective hypergraph to a bond in the lattice similar to the description for conventional graphs in \cite{oitmaa_hamer_zheng_2006}. \par 

As this requirement is considered during isomorphism checking, we will continue with the sorting of the instances into isomorphism classes. To sort a subcluster into an isomorphism class, we use its reduced K\"onig representation as described in Sec.
\ref{subsect::sorting}.
We first calculate a combination of heuristic graph invariants, and then perform isomorphism checks using the RI algorithm \cite{Bonnici2013subgraph, Bonnici2016variable} among the already discovered graphs with the same invariants.
If no isomorphic graph is found, we discovered a new isomorphism class, which we insert into the list of isomorphism classes with an embedding count of one.
If an isomorphic graph is found, we check whether there is an isomorphism which preserves the first vertex,~i.e., it maps the first vertices of both graphs onto each other. If such an isomorphism exists we increase the embedding count of the respective equivalence class by one.
If no such isomorphism exists, the current instance is associated with another bond and the embedding count is not increased. In this way, we define a canonical position for the first vertex for every isomorphism class which for example can be based on the first obtained representant. Whenever we accept a subcluster based on the previous criteria, we increase the embedding count by one.
If the canonical position of the first vertex is ambiguous, i.e. the corresponding automorphism orbit contains multiple elements, we need to correct the final embedding number dividing it by the size of the respective orbit.
In this way we find a list of subclusters and their embedding numbers and can proceed to 
the final evaluation of matrix elements, which can be done
with the commonly employed approaches like bookkeeping techniques \cite{coester2015optimizing} or recursive subcluster subtraction schemes \cite{Gelfand2000, oitmaa_hamer_zheng_2006}.

\subsubsection{Vacuum energy series} 
We calculated the vacuum energy per plaquette for \eqref{eq::application} up to order twelve
\begin{equation}
\begin{aligned}
e_0 = &-\frac{1}{3}-\frac{{{\lambda}^{2}}}{8}-\frac{113 {{\lambda}^{4}}}{1536}-\frac{21427 {{\lambda}^{6}}}{163840}-\frac{87959384893 {{\lambda}^{8}}}{254803968000}\\&-\frac{115181804621864639 {{\lambda}^{10}}}{102736959897600000} \\&-\frac{1199864820008961969940451  {{\lambda}^{12}}}{289964795614986240000000} \, ,
\end{aligned}
\end{equation}
using L{\"o}wdin's partitioning technique \cite{Loewdin1962, Kalis2012} with a bookkeeping technique \cite{coester2015optimizing} for the evaluation. After proper rearrangements this agrees with the available result in order six obtained in \cite{Muehlhauser2020}. We have therefore significantly extended the maximal order of the vacuum energy series with our hypergraph approach. 
 
 \FloatBarrier
 
%%%%%%%%%%%%%%%%%%%%%%%%%%%%%%%%%%%%%%%%%%%%%%%%%%%%%%%%%%%%%%%%%% 
\subsection{Hopping elements}
%%%%%%%%%%%%%%%%%%%%%%%%%%%%%%%%%%%%%%%%%%%%%%%%%%%%%%%%%%%%%%%%%%

The calculation of hopping elements is quite analogue to the previously described calculation of the vacuum energy. The main difference is that we now need all subclusters which contain at least one site which is occupied, i.e., different from the local vacuum, in the initial state. Furthermore any considered cluster should contain all sites, which change their local state in the matrix element of interest.
Accordingly, we start the search process several times, from all the bonds containing any of the sites which are occupied in the initial state. Obviously, when we restart the search from a different bond, all the previous starting bonds are ignored in order to avoid duplicates as described in \cite{Ruecker_2000}. \par 
We also have to slightly modify the employed equivalence relation, as the equivalence of subclusters now depends crucially on the states involved in the matrix element.
To this end we distinguish the vertices based on the associated local states. 
In our case it is sufficient to distinguish between four possibilities, namely occupied in the initial state, occupied in the final state, both or none of the two.
As a consequence, we have to slightly tweak the reduced representation which we used for the vacuum energy. This can be achieved by keeping the vertices of degree one which correspond to sites occupied in the final state, in the initial state, or in both states.
In this setting simply testing for isomorphism is sufficient, and we do not need the first vertex to be in a canonical location as for the vacuum energy. The embedding counts of the isomorphism classes are simply given by the number of subclusters which belong to the given class. \par
Also the heuristics to discard non-contributing clusters or equivalence classes have to be adapted.
This can be done by assuming that only sites which are unoccupied in the initial and final state require to be modified an even number of times by the perturbation.
Depending on which technique is employed only these sites or vertices have to be counted or covered.
For the problem at hand we can also require that sites which remain occupied have to be affected an even number of times by the perturbation, and sites which change their occupation number are touched an odd number of times. 
In addition we can employ another check, which determines whether all sites which have to be on the subcluster can be reached by expanding the subcluster up to a given number of bonds. If this is not the case the subcluster is not further expanded and discarded. The distances from the relevant sites can be calculated once before the actual search processes start. \par
 To give an exemplary result we calculated the gap of two adjacent spin-flip excitations for the transverse field plaquette Ising model \eqref{eq::application}, as this is the energetically lowest mobile excitation \cite{Shirley2019, Muehlhauser2020}.
Observing that the parity of the excitation number in each plane is conserved, one sees that single spin-flip
excitations are immobile \cite{Vijay2016}, and the described two spin-flip excitations stay together as nearest neighbours with fixed orientation and can only move within a plane \cite{Slagle2018}. As a consequence, this particle sector can be diagonalized using Fourier transformation. The gap $\Delta_2$ of the resulting dispersion is located at $\vec{k}=0$ and reads

\begin{equation} 
\begin{aligned}
\Delta_2 = &4-4 \lambda-3 {{\lambda}^{2}}-\frac{17 {{\lambda}^{3}}}{8}-\frac{1151 {{\lambda}^{4}}}{192}-\frac{37165 {{\lambda}^{5}}}{9216} \\ &-\frac{2591423 {{\lambda}^{6}}}{122880}-\frac{6264944713 {{\lambda}^{7}}}{530841600} \\
&-\frac{5707377242657 {{\lambda}^{8}}}{63700992000}-\frac{2114517232207 {{\lambda}^{9}}}{53084160000} \\ &-\frac{2169534326790862117 {{\lambda}^{10}}}{5136847994880000}\, , \end{aligned} \end{equation}

which coincides with the series up to order seven obtained in \cite{Muehlhauser2020}. Here we evaluated the contributions of the subclusters using pCUT with a bookkeeping technique \cite{coester2015optimizing}. Again, our approach provides a significant extension of the maximal order (see also App.~\ref{appendix::calc}).\par

%%%%%%%%%%%%%%%%%%%%%%%%%%%%%%%%%%%%%%%%%%%%%%%%%%%%%%%%%%%%%%%%%%%%%%%%%%%%%%%%%%%%%%%%%%%%%%%%%%%%%%%%%%%%%%
\section{Conclusion}
\label{sect::conclusion}
%%%%%%%%%%%%%%%%%%%%%%%%%%%%%%%%%%%%%%%%%%%%%%%%%%%%%%%%%%%%%%%%%%%%%%%%%%%%%%%%%%%%%%%%%%%%%%%%%%%%%%%%%%%%%%
In this work we presented a straightforward and relatively general approach to incorporate effective many-site interactions into graph decompositions. Although the involved methods and theoretical foundations are well established, we believe that the presented combination of all this to a general approach for multi-site interactions is valuable for perturbative linked cluster expansions of quantum many-body systems.

It is clear that the presented scheme cannot compete with the well established graph decomposition techniques \cite{Gelfand2000, oitmaa_hamer_zheng_2006} where they are applicable. One of the reasons is the large overhead of the internal representation of hypergraphs and related information. 
%\mmnew{Obviously, also our implementation leaves a lot of space for improvements, e.g., we used the RI-algorithm \cite{Bonnici2013subgraph} for graph isomorphism, which is originally designed for subgraph-isomorphism.
%It would be interesting to investigate the impact of state of the art canonical labeling algorithms like nauty or traces \cite{McKay2014}.} \mmcomment{Do we need this?} 
Furthermore, our approach to identify non-contributing clusters can still be improved by using more efficient algorithms for subtasks, like the partitioning of the set of odd-degree vertices into bonds, but also by conceptional improvements of the heuristics themselves. 

%From a more general perspective we hope that this method helps to access more involved problems, like
%the calculation of a high-field expansion for the X-Cube model in an arbitrary magnetic field analogously to the 2D-toric code \cite{kamfor2013robustness}.
%\mmcomment{Finde ich bisschen schwierig weil das Bottleneck variiert. Im Vacuum ist es z.B. mittlerweile die DT filterung, aber ohne DT Filterung macht isomorphismen checken doch noch viel aus}
As quantum many-body systems containing multi-site interactions play an important role in current research due to their exotic physical properties like topological or fracton order, we are strongly convinced that the presented approach can be successfully used in various systems in the future. It would be further interesting to use the described hypergraph approach with non-perturbative numerical linked cluster expansions.

\begin{acknowledgments}
	MM would like to thank his colleagues, especially Carolin Boos and Max H{\"o}rmann, for critical and fruitful discussions.
	KPS acknowledges financial support by the German Science Foundation (DFG) through the grant SCHM 2511/11-1.

\end{acknowledgments}

\begin{appendix}

	\section{More general settings}
		\label{appendix::complicatedmodels}
		In this appendix we discuss relatively common generalizations and propose solutions.
		The first generalization is the presence of several bond types. While for the plaquette Ising model in a transverse field \eqref{eq::application} we were able to
		directly calculate the vacuum energy per plaquette, the presence of different bond types requires some
		further considerations. 
		Obviously, the graph representation is modified taking into account the bond type as a vertex label assigned to the vertices in the edge-partition, which should at the same time still be distinguishable from the vertices in the vertex-partition. 
		For the vacuum energy one has to be careful ensuring the right embedding counts. One possibility is to define some sort of unit cell, where all bond types are contained and which is periodically repeated. 
		If all bond types in this unit cell are different, we can simply start the search process from every bond type ignoring the bond types of the former starting bonds, as all relevant subclusters involving former starting bonds have already been discovered.
		In principle if some bond type appears multiple times in such a unit cell, one can just distinguish bond types arbitrarily. 
		For the calculations of the hopping elements we do not expect such complications, and thus 
		assigning additional vertex labels for the edge-partition should suffice. \par
		The treatment of more than two local states should only require modifications for the calculation of hopping elements, as more local states have to be distinguished. 
		\par
		To treat directed interactions, where the local interaction matrix elements depend on the specific sites within a bond, we propose to use an edge-colored version of the reduced K{\"o}nig representation in order to distinguish the different sites in the bond.
		Actually, this can be interpreted as the representation of a sequence hypergraph, i.e. a hypergraph where the hyperedges are given by sequences instead of sets \cite{Boehmova2016}.
		Although they can be seen as a generalization of directed graphs, they are not to be confused with directed hypergraphs \cite{Boehmova2016, Gallo1993, Bretto2013} 
						
%	\section{Parallelization}
%		
%			The presented graph decomposition technique is parallelizable as several branches %of the search-tree can be transversed simultaneously, keeping a local list of equivalence %classes for every branch. However, especially for the maintaining of canonical orbits we still %require some communication between the threads.
%			Furthermore, in the end of the calculation the branches have to be merged to get %the overall embedding counts. 
%			We implemented a shared memory parallelization using openmp, and determined the %speed-up for several numbers of threads as given in Fig. \ref{Amdahl1}. 
%			\begin{figure}[htb]
%				\includegraphics[width=0.5 \textwidth]{Figures/AmdahlForPaper.pdf}
%				\caption{Speed-up over number of threads for the calculation of graphs which we %used to obtain the vacuum energy. The scaling is not good, but for relatively low numbers of %cores the parallelization still provides a good speedup. We also fitted Amdahl's law %\cite{Amdahl67, Hill2008multicore} to the data.}	
	%			\label{Amdahl1}  
	%		\end{figure}
	%		We observe that we get good speed-ups only for low numbers of threads. 
	%		For the single threaded version we observed a run, which took roughly twice the 
	% CPU-time of the other single threaded runs. 
	
	\section{Truncating the search tree using graph coloring}
	\label{appendix::graphcoloring}
\begin{figure}[htbp]
		\includegraphics[width = 0.5 \textwidth]{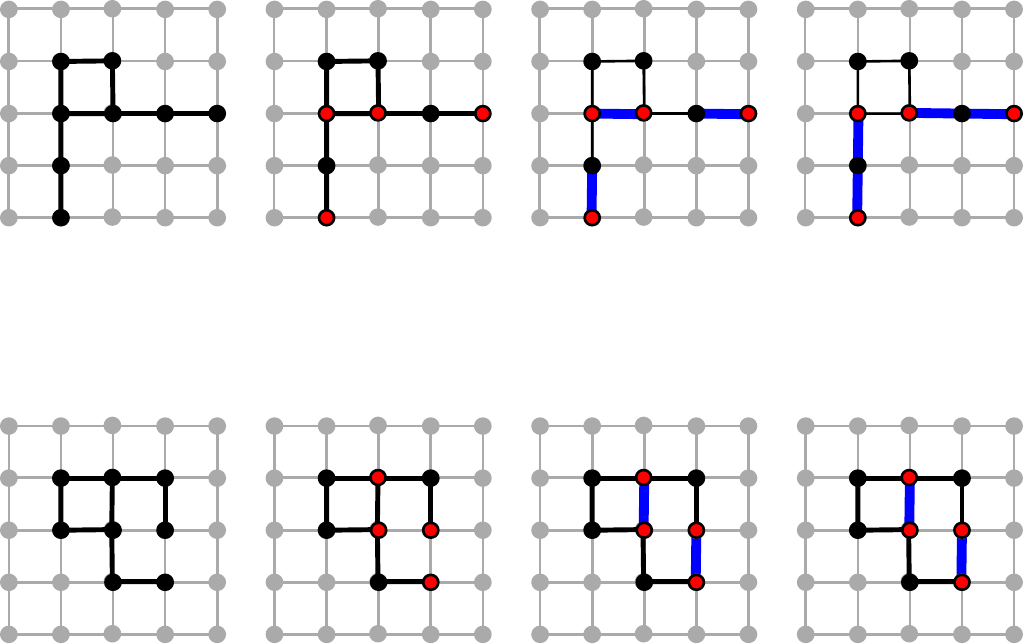}
		\caption{Considerations about two different clusters (upper/lower row) which belong to the same isomorphism class. From left to right: We first give the position of the cluster in the actual lattice. Then in the next column we mark all vertices of odd-degree red. In the third column we mark a minimum set of bonds, such that all odd-degree vertices are covered. To show that this is again only an estimate, we consider coverings, where all vertices in the cluster will have even degree after addition (duplication) of the blue edges.
			\label{fig::embedding}}
	\end{figure}

	In this appendix we aim to give a simple example for the transitive heuristic to discard non-contributing clusters for which we use graph coloring. 
	To this end we consider a calculation of the ground-state energy of the well known transverse field Ising model on a square lattice in the high-field limit
	\[ H^{\lambda \ll 1}_{\text{TFIM}} = -\sum_i \sigma_i^z - \lambda \sum_{\langle i,j\rangle} \sigma^x_i \sigma^x_j \qquad \lambda \ll 1 \, .\]
	We consider this example, because it allows us to keep images comprehensive and simple, and it is readily generalized to other suitable problems.
	In Fig.~\ref{fig::embedding} we show the embeddings of two clusters which one encounters within the calculation of the vacuum energy and belong to the same isomorphism class.
	For both of these embeddings we go through four steps, which are represented by the rows in Fig.~\ref{fig::embedding}.
	The first step is to specify the actual realization of the cluster within the lattice.
	Then we identify the sites, which have odd-degree in each cluster. As described in Sec.~\ref{subsubsect::non-contributing} this information
	can already be used to discard non-contributing clusters \cite{oitmaa_hamer_zheng_2006}. In the present case, the corresponding graph has 8 edges and 4 vertices of odd degree. As each bond covers 2 vertices, we can conclude, that at least two edges have to be added to get rid of the odd-degree vertices. Accordingly, this graph can contribute only in order ten or higher. \par
	In an actual calculation we can store the adjacency information of all the sites in the lattice, i.e.~we store whether two sites share a common bond.
	We can use this information to find a lower bound to the number of bonds which are needed in order to cover the sites of odd degree. Obviously, two sites, which do not share a bond, cannot be covered by the same bond. So we search for a partition of the odd-degree sites, such that non-adjacent sites are in different parts. A lower bound to the number of parts is actually given by the chromatic number.
	The chromatic number of a graph is the minimum number of colors necessary to find a feasible coloring,~i.e. a coloring which assigns a color to every vertex of the graph, such that adjacent vertices never have the same color \cite{Lewis2016}.
	The chromatic number is calculated for the graph where vertices represent odd-degree sites and pairs of vertices are joined by an edge, whenever they do not share a bond. It is intuitively clear that pairs of vertices, which do not share a bond, will be assigned different colors. And thus the coloring provides exactly the partition we searched for. 
	\begin{figure}[htbp]
		\includegraphics[width = 0.1 \textwidth]{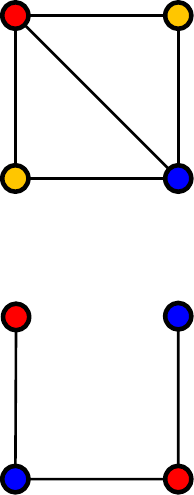}
		\caption{The graphs which have to be colored to check the two clusters in Fig.~\ref{fig::embedding}.
		Due to the different embeddings in the lattice of the corresponding clusters the upper graph has a higher chromatic number, and thus we need more bonds to cover the odd-degree sites. 
		}
	\end{figure}
	Note that using graph coloring, to solve these kinds of problems is a standard technique \cite{Lewis2016}. 
	Of course, when checking if we can discard a cluster and its descendants we do not need to know the chromatic number. We just need to know whether it exceeds the number of bonds which can be appended down the search tree. So instead of computing the chromatic number, we check if it is possible to find a feasible coloring with $r$ colors, where $r = (O_\text{pert}  -  N_{\text{bonds}})$.
	This can be decided with a simple backtracking algorithm. As this decision is even simpler for $r=1,2$ specialized solutions for these cases should be considered for optimization.
	We also see that these heuristics can still be improved, as duplicating the three blue bonds on the cluster in the upper row of Fig. \ref{fig::embedding} does create new odd-degree sites. \par
	In the fourth step, we provide a set of bonds which does not leave any odd-degree sites, in order to show, that there is still a lot of space for improvement in the pruning of the search tree for these kinds of models.

	\section{Comparison to calculations on single clusters}
	\label{appendix::calc}
	In order to estimate the benefit of the presented linked cluster expansions compared to
	calculations on sufficiently large clusters in the context of the examples given in Sec.~\ref{sect::application}, we performed two explicit calculations collecting some information about CPU-time and memory usage with the help of GNU time. \par	

	First we determined the vacuum energy per plaquette up to order 6 evaluating the effective pCUT-Hamiltonian on a periodic cluster of $7\times7\times7$ spins. This calculation took more than 30 minutes of CPU time, and over 30 GB of maximum resident set size (RSS). Instead, with the described method we obtained equivalent results in less than 1 s of CPU-time and a maximum RSS below 5 MB on the same system.
	Furthermore, we calculated the matrix element of two adjacent spin excitations hopping by one site on a cluster with $8\times8\times7$ spins. The calculation took more than 25 hours of CPU time and a maximal RSS of around 70 GB, while with the presented method it took less than 4 minutes of CPU time and a maximal RSS of less than 15 MB to obtain the same result.
	For both of these calculations pCUT with a bookkeeping technique \cite{coester2015optimizing} was used for evaluation.

\end{appendix}
\newpage
\bibliography{References.bib}

\end{document}